\documentclass{aa}

\usepackage{stfloats}
\usepackage{graphicx}
\usepackage{multirow}
\usepackage{txfonts}
\usepackage[linkcolor = blue]{hyperref} 
\usepackage{xcolor}
\usepackage{placeins}
\usepackage{soul}

\makeatletter
\renewcommand*\aa@pageof{, page \thepage{} of \pageref*{LastPage}}
\makeatother

\usepackage{natbib}
\bibpunct{(}{)}{;}{a}{}{,} 

\begin{document}

\title{Multi-mode Pulsations in AGB Stars:\\ Insights from 3D RHD {CO5BOLD} Simulations}



   \author{A. Ahmad,
          B. Freytag
          \and
          S. Höfner
          }

   \institute{Theoretical Astrophysics, Division for Astronomy and Space Physics, Department of Physics and Astronomy, Uppsala University,
                Box 516, 751 20 Uppsala, Sweden\\
                \email{arief.ahmad@physics.uu.se}
         }

   \date{Received February 17, 2025; accepted MONTH DAY, YEAR}


  \abstract
   {Stars on the asymptotic giant branch (AGB) can exhibit acoustic pulsation modes of different radial orders, along with non-radial modes, throughout their evolution. These pulsations are essential to the mass-loss process and influence the evolutionary pathways of AGB stars. Period–luminosity (P–L) relations serve as a valuable diagnostic for understanding stellar evolution along the AGB. Three-dimensional (3D) radiation-hydrodynamic (RHD) simulations provide a powerful tool for investigating pulsation phenomena driven by convective processes and their non-linear coupling with stellar oscillations.}
   {We investigate multi-mode pulsations in AGB stars using advanced 3D ‘star-in-a-box' simulations with the \texttt{CO5BOLD} RHD code. Signatures of these multi-mode pulsations were weak in our previous 3D models. Our focus is on identifying and characterising the various pulsation modes, examining their persistence and transitions, and comparing the results with one-dimensional (1D) model predictions and observational data where applicable.}
   {We produced a new model grid comprising AGB stars with current masses of $0.7$, $0.8$, and $1\,\mathrm{M}_{\odot}$. Fourier analysis was applied to dynamic, time-dependent quantities to extract dominant pulsation modes and their corresponding periods. Additionally, wavelet transforms were employed to identify mode-switching behaviour over time.}
   {The simulations reveal radial, non-radial, fundamental, and overtone modes, with their transitions and dominance depending on stellar parameters. The models successfully reproduce the P–L sequences found in AGB stars. Mode-switching phenomena are found in both the models and wavelet analyses of observational data, allowing us to infer similarities in the underlying pulsation dynamics. The results confirm the dependence of pulsation periods on mean stellar density and underscore the significant role of convection for the amplitude of multi-mode pulsations.}
   {These 3D simulations highlight the natural emergence of multi-mode pulsations, including both radial and non-radial modes, driven by the self-consistent interplay of convection and oscillations. Our findings underscore the value of 3D RHD models in capturing the non-linear behaviour of AGB pulsations, providing insights into mode switching, envelope structures, and potential links to episodic mass-loss events.}


   \maketitle
%

\section{Introduction}

The pulsation characteristics of evolved stars, particularly asymptotic giant branch (AGB) stars and red supergiants (RSGs), provide valuable insights into their internal structure, evolutionary processes, and mass-loss mechanisms. Stars along the AGB, which originate as low- to intermediate-mass stars during their main-sequence phase, enter a brief yet critical late evolutionary stage where pulsations play a key role in driving mass loss. This mass loss subsequently enriches the interstellar medium with metals in the form of gas and dust. Despite their short-lived nature compared to the core hydrogen and helium burning phases, the contribution of AGB stars to the chemical evolution of galaxies is substantial, making their study essential \citep[e.g.][]{tosiAGBStarsChemical2007, karakasHeavyelementYieldsAbundances2018}.

In AGB stars, pulsations arise from the interplay between stellar convection and oscillatory acoustic modes, generating shock waves that propagate through the atmosphere. These pulsation-driven shocks lift material to cooler regions where dust formation becomes favourable, enabling radiation pressure to accelerate stellar winds \citep[see, e.g.][for a recent review]{hofnerMassLossStars2018}. 
Consequently, the atmospheric dynamics give rise to intricate circumstellar structures, such as pronounced radial layers \citep[e.g.][]{fleischerCircumstellarDustShells1992a, bladhExtensiveGridDARWIN2019a}, with the resulting velocity variations affecting near-IR line profiles \citep[e.g.][]{nowotnyLineFormationAGB2010a, liljegrenPulsationinducedAtmosphericDynamics2017b}, and clumpy wind regions \citep{freytagGlobal3DRadiationhydrodynamical2023}. Observations reveal that AGB stars exhibit spectral energy distributions shifted towards the infrared, which is a signature of dust reprocessing the stellar radiation. The three-dimensional (3D) morphology of circumstellar dust clouds will affect both spatially resolved observations and spectral energy distributions of unresolved objects \citep{wiegertAsymmetriesAsymptoticGiant2024}. Therefore, understanding pulsation behaviour is essential for elucidating the formation of these structures and their role in stellar feedback processes \citep{piovanShellsDustAGB2003, maerckerDetachedDustShells2014, villaumeCircumstellarDustAGB2015}.

One of the central challenges in modelling the pulsations of AGB stars lies in accurately representing the interaction between convection and pulsation modes. Traditional one-dimensional (1D) codes assume spherical symmetry and apply a parameterised treatment of convection, which can oversimplify the dynamic processes that occur in AGB stars. In contrast, 3D radiation-hydrodynamics (RHD) simulations have shown that convection and pulsations are deeply interconnected, generating large-scale shocks in the stellar atmospheres of AGB stars \citep{freytagGlobal3DRadiationhydrodynamics2017, freytagGlobal3DRadiationhydrodynamical2023}.

Recent advancements in observations have provided further validation of multi-mode pulsations in AGB stars. Ground-based surveys such as OGLE and space missions like \textit{Gaia} and \textit{Kepler} have confirmed the presence of fundamental and overtone pulsation modes across different evolutionary stages \citep{mosserPeriodluminosityRelationsEvolved2013, yuAsteroseismologyLuminousRed2020, iwanekOGLECollectionVariable2022}. These observations have refined the period-luminosity (P-L) relations, revealing distinct P-L sequences associated with multi-mode pulsations \citep{woodMACHOObservationsLMC1999b}. Generally, red giants initially pulsate in low-amplitude non-radial modes. As they evolve, expand, and cool, their pulsations transition to radial modes with longer periods and larger amplitudes, eventually favouring the fundamental mode. The validation of P-L relations using 1D models \citep{trabucchiModellingLongperiodVariables2019} has provided essential constraints on pulsation properties and stellar parameters, demonstrating the predictive power of theoretical models when compared with observations.

Mode switching, where a star changes its dominant pulsation mode (defined as the mode with the highest amplitude), is another phenomenon closely tied to stellar evolution. This transition is expected as stars evolve across different P-L sequences, with observational evidence showing that red giants can switch between fundamental and overtone modes \citep{kissMultiperiodicitySemiregularVariables2000a, hinkleVelocityObservationsMultipleMode2002}. 
The combined amplitude of decaying higher-overtone modes and emerging lower-overtone modes may be a key factor in triggering more complex pulsation behaviours, such as the long secondary period \citep{pawlakConnectionLongSecondary2021}. The transition from small-amplitude red giants to semiregular variables (SRVs), associated with the onset of self-excited pulsations, is linked to changes in the stellar envelope \citep{trabucchiSelfexcitedPulsationsInstability2025}. These changes in the pulsation modes are influenced by factors such as metallicity, hydrogen content, and the treatment of turbulent viscosity in the models. The sensitivity of the transitions of the pulsation modes to the stellar envelope composition provides critical constraints for theoretical models, particularly in understanding the evolutionary pathways of long-period variables (LPVs) \citep{trabucchiSelfexcitedPulsationsInstability2025}. Mode switching offers insights into the internal structure of stars and the mechanisms governing the stability of pulsation modes, with implications for stellar feedback and the calibration of P-L relations.

The interaction between pulsations and convection not only influences internal structures but also surface phenomena such as granulation. Surface patterns, closely tied to surface gravity, are shaped by convection \citep{freytagGlobal3DRadiationhydrodynamics2017, freytagDimmingEventsEvolved2024} and the effects of pulsations \citep{rosales-guzmanNewDimensionVariability2024}. High-resolution observations, such as infrared interferometry, have revealed large-scale convection cells and surface inhomogeneity in evolved stars \citep{paladiniLargeGranulationCells2018, ohnakaInfraredInterferometricThreedimensional2019}. Multi-mode pulsations, particularly non-radial modes, can further complicate these surface patterns, leading to irregular light curves and variability, as observed in LPVs and SRVs \citep{kissMultiperiodicitySemiregularVariables1999b, cadmusLongtermPhotometricBehavior2024}. These surface variations influence observational diagnostics of wind properties by contributing to the formation of asymmetrical dusty structures \citep{wiegertAsymmetriesAsymptoticGiant2024}.

While significant progress has been made, challenges remain in studying overtone modes and mode dominance using 3D simulations. Despite the complex, non-linear interactions between convection and pulsations, the pulsation periods of AGB stars correlate well with their mean density, which is influenced by fundamental stellar parameters \citep{ahmadPropertiesSelfexcitedPulsations2023a}. Previous 3D studies of AGB stars have primarily focused on the radial fundamental mode, as overtone signatures have remained too weak to allow detailed investigation of mode-switching behaviour and energy redistribution. Given the pivotal role of large convective envelopes, such as with evolved cool giants, to investigate amplitude variations, phase shifts, and pulsation mode stability, further advancements in simulation techniques are essential to capture these complex non-linear processes fully \citep{ freytagGlobal3DRadiationhydrodynamics2017}. 

In this work, we explore the intricate nature of multi-mode pulsations in AGB stars using a new grid of global 3D RHD models produced with the \texttt{CO5BOLD} code. Our analysis focuses on identifying dominant pulsation modes, investigating the conditions for mode switching, and examining the interplay between pulsations, convection and atmospheric dynamics. By combining observational insights with 3D simulations, we aim to contribute to a broader understanding of stellar variability, mode evolution, and the complex processes shaping the late stages of stellar evolution along the AGB.

\section{Method}

\subsection{Setup of 3D models}\label{sect_method_3dmodels}

For our simulations, we used the \texttt{CO5BOLD} code \citep{freytagSimulationsStellarConvection2012a, freytagAdvancesHydrodynamicsSolver2013, freytagBoundaryConditionsCO5BOLD2017}, which numerically integrates the coupled non-linear equations of hydrodynamics and radiation transport. The compressible hydrodynamics equations enable the modelling of travelling pressure waves, standing acoustic modes, transonic convective flows within the stellar interior, and shocks in the atmosphere. Using the 3D RHD \texttt{CO5BOLD} code, we have generated a new grid of AGB models. A more detailed description of \texttt{CO5BOLD}, particularly as it relates to the work presented here, can be found in \cite{ahmadPropertiesSelfexcitedPulsations2023a}.

\subsection{Advances over previous generation of models}\label{sect_method_differencesinmodels}

The models analysed in \cite{ahmadPropertiesSelfexcitedPulsations2023a}
had a range of masses and luminosities covering AGB to RSG stars.
But all of them had rather low mean densities, favouring fundamental mode pulsations.
The simulations were performed over several years with different versions of CO5BOLD
and a variety of numerical setups.
In contrast, our new grid of models of AGB stars,
covering a small range of masses but a large span in luminosities,
keeps the numerical parameters consistent.
A key advancement is the focus on exploring denser stars along the AGB. 
We have produced a grid of models with current masses of 0.7, 0.8, and 1.0 $\mathrm{M}_{\odot}$, 
all with a core mass of $0.6 \, \mathrm{M}_{\odot}$. This setup is guided by core mass predictions 
derived from 1D stellar evolution models, which indicate this value is representative of late AGB 
evolutionary stages \citep[e.g.][]{karakasHeavyelementYieldsAbundances2018, venturaGasDustExtremely2021, cinquegranaBridgingGapIntermediate2022}.

A significant improvement in this new grid is the adjustment and consistent application of the 
gravitational smoothing parameter, $r_{0}$ \citep[see][]{ahmadPropertiesSelfexcitedPulsations2023a}, 
across the new model grid. This was set to $7 \, \%$ of the stellar radius ($\mathrm{R}_{\star}$), in 
comparison to between $20-25\,\%$ of $\mathrm{R}_{\star}$ in our previous models. We define 
$\mathrm{R}_{\star}$, as the point of minimum entropy, based on prior investigations into the 
optimal definition of this parameter in our models \citep{ahmadPropertiesSelfexcitedPulsations2023a}. 
The $r_{0}$ parameter prescribes the smoothing of the stellar gravitational potential to account for 
the unresolved compact core, where most of the stellar mass resides, given that we have set the core 
mass of $0.6 \, \mathrm{M}_{\odot}$. Within $2 \, r_{0}$, energy is injected at a constant rate
according to the desired luminosity.
The size of the computational domain and the number of grid points were chosen to
keep the number of cells per surface pressure height similar
and to account for the different extensions of the atmospheres of models with different surface gravities.

We designate three representative models in this article to discuss specific methods and results. All three have a current stellar mass of $1\,\mathrm{M}_{\odot}$. Their fundamental parameters are listed in Table~\ref{table:model_representatives}. Chosen to illustrate distinct self-excited pulsation behaviours, the models are referred to by the following letters: a Mira-like star (model \texttt{A}), a star with irregular pulsations (model \texttt{B}), and a star with strong overtone pulsations (model \texttt{C}).

\begin{table}[t]
\caption{Parameters of representative stellar models.}  
\label{table:model_representatives}
\centering
\begin{tabular}{c c c c c}
\hline\hline                                   
Model & Designation & $R_\star\,[R_\odot]$ & $T_\text{eff} \, [\text{K}]$ & $L_\star \, [L_\odot]$ \\                     
\hline     
st28gm06n073 & \texttt{A} & 321 & 2824 & 5914 \\
st28gm05n053 & \texttt{B} & 293 & 2846 & 5088 \\
st31gm01n003 & \texttt{C} & 167 & 3135 & 2436 \\

\hline                                                     
\end{tabular}
\tablefoot{All three models are of current stellar mass $1\,M_\odot$, and the stellar parameters are temporally-averaged values.}
\end{table}

\subsection{Pulsation period extraction process}\label{sect_method_pulsation_extraction}

The extraction of pulsation periods begins with the application of the Fast Fourier Transform (FFT) to temporal quantities selected according to the specific characteristics of the pulsation modes under investigation. Unless stated otherwise, all references to power spectra in this article pertain to those derived from the FFT. To analyse purely radial motions, the radial mass flux and mass density are averaged over spherical shells. These spherically averaged quantities are denoted as <.>$_{\Omega}$, where $\Omega$ represents the solid angle of a sphere. The ratio of the averaged radial mass flux, <$\rho v_{\textrm{radial}}$>$_{\Omega}(r,t)$ to the averaged mass density, <$\rho$>$_{\Omega}(r,t)$, yields the radial velocity as a function of radial distance and time, $v_{\textrm{radial}}(r,t)$. 
We also introduce the spherically averaged quantity $\sqrt{\rho}v_{\mathrm{radial}}(r,t)=\textrm{<}\rho v_{\textrm{radial}}\textrm{>}_{\Omega}(r,t)/\textrm{<}\rho\textrm{>}_{\Omega}^{0.5}$$(r,t)$, which is used in Sect.~\ref{sect_method_nodes}.
Both spherical‐averaging procedures follow the methodology used in \citet{freytagGlobal3DRadiationhydrodynamics2017} and \citet{ahmadPropertiesSelfexcitedPulsations2023a}. To account for non-radial pulsation modes, this method is extended using spherical harmonics. By incorporating weights associated with spherical harmonics into the spherical averaging, where each degree is denoted with $\ell$, there are $2\ell+1$ spherical harmonics ($-\ell \le m \le +\ell$); we obtain the velocity as a function of $\ell$ and $m$. For each degree, we sum the power spectra across each possible $m$ to obtain a single power spectrum representative of each $\ell$.

\begin{figure}[t]
   \centering
   \includegraphics[width=\hsize]{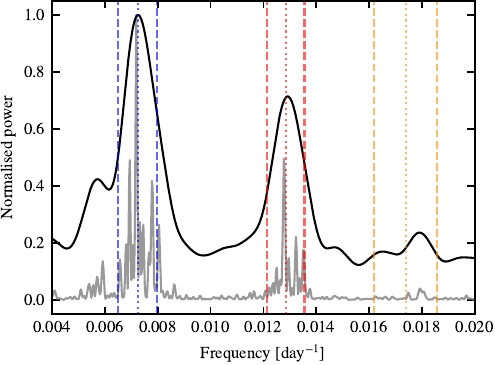}
      \caption{The grey curve shows the averaged power spectrum (normalised so that its maximum is unity) of the spherically averaged radial velocity, computed over the radial range $80-100\, \%$ of $\mathrm{R}_{\star}$, for model \texttt{C}. The black curve shows the same signal after applying the two-step smoothing procedure (Savitzky–Golay followed by Gaussian filtering), demonstrating the effectiveness of the method in reducing noise while preserving peak structure. Dotted and dashed vertical lines indicate the derived $f_\text{mean}$ values and their corresponding bounds $f_{\textrm{mean}} \pm f_{\textrm{RMS}}$ for the fundamental, first overtone, and second overtone modes, shown in blue, red, and orange, respectively. See main text for further details.}
         \label{fig_fftprocedure}
   \end{figure}

Applying the FFT to the spherically averaged velocity yields FFT spectra as a function of radial distance from the stellar centre, corresponding to layers within the stellar interior. To account for spatial variability across these radial layers, an average power spectrum is computed for the region spanning $80-100\, \%$ of $\mathrm{R}_{\star}$. This region consistently exhibits clear signals of pulsations and their effects \citep[see][]{ahmadPropertiesSelfexcitedPulsations2023a}, and the resulting averaged power spectrum is thus used as the representative signal for the subsequent analysis. To enhance the clarity of this signal, a combination of sequential smoothing methods is applied. 
First, a Savitzky-Golay filter is applied to the spectrum using a fourth-degree polynomial and a symmetric window defined by a cut-off frequency of $10^{-3}$ cycles per day. This step retains local features while removing small-scale fluctuations. Second, the output is convolved with a Gaussian kernel with $\sigma=5$ frequency bins (i.e. five times the frequency resolution of the power spectrum) to suppress residual noise. 
This combined approach has proven effective in retaining the primary pulsation features by preserving peak structures while minimising spurious variability. The improvement is illustrated in Fig.~\ref{fig_fftprocedure}, where the final smoothed signal more clearly reveals the local minima surrounding each peak. This facilitates the identification of well-defined frequency bounds for regions containing significant signal power.
Overall, this ensured reliable identification of pulsation features in Fourier space, and mitigates non-linear effects such as interference by convection.

%

Significant frequencies are identified using an iterative peak-detection procedure inspired by pre-whitening techniques in asteroseismology. At each iteration, the strongest peak in the current power spectrum is located, and an initial frequency range encompassing the peak is defined, by bounding the peak with the nearest local minima on either side of the peak. This region is then refined by computing the weighted mean frequency ($f_{\textrm{mean}}$) and weighted root mean square (RMS) values ($f_{\textrm{RMS}}$) using the power at each frequency bin as weights. We take the frequency region $f_{\textrm{mean}} \pm f_{\textrm{RMS}}$ to be associated with the present pulsation mode. The power contained within this region is normalised by the total spectral power of the original signal, yielding a fractional power that serves as a measure of the mode's significance. Once characterised, the region is removed (i.e. zeroed) from the original spectrum, and the process is repeated on the residual, including the two-step smoothening process. 

Using the procedure described in the preceding paragraph, we identify up to 20 regions. If any of the regions overlap, that is, if their intervals (expanded by the constant factor 1.1 to $f_{\textrm{mean}} \pm 1.1f_{\textrm{RMS}}$ ) intersect, they are merged into a single region.
The new bounds are set to the minimum and maximum frequencies across the overlapping intervals. Within this merged region, the $f_{\textrm{mean}}$, $ f_{\textrm{RMS}}$ and power fraction are recomputed using the updated window.
A justification for this step is seen in Fig.~\ref{fig_fftprocedure}: the weaker signal just below $0.006\,\textrm{day}^{-1}$ should be considered as part of the fundamental mode \citep[attributed to non-linear effects, as discussed in][]{ahmadPropertiesSelfexcitedPulsations2023a}. But this region would initially be identified as separate from the stronger signal between $0.006-0.008\,\textrm{day}^{-1}$. Merging consolidates these two intervals into a single mode.
At the end of this procedure, a minimum of one frequency window remains which correspond to the fundamental mode, and up to four windows may be identified in total, depending on the stellar parameters. These additional windows are associated with overtone modes as seen in Fig.~\ref{fig_fftprocedure} and are discussed further in Sect.~\ref{resultssection}. 
For each window, the pulsation period and its spread (could be accounted for as its uncertainty) is then calculated with $P_{\textrm{puls}}=1/f_{\textrm{mean}}$, and its spread $\sigma_{\textrm{puls}}=P_{\textrm{puls}} (f_{\textrm{RMS}}/f_{\textrm{mean}})$.

The method outlined in this section successfully identifies various pulsation modes across different radial orders. The overtone order is determined systematically: the fundamental mode corresponds to the lowest frequency, the first overtone to the second-lowest, and so on. Non-radial modes are distinguished by the velocity components used in the power spectrum analysis. The derived pulsation frequencies and their spreads are used to calculate the fraction of power within each mode's frequency band. Consequently, we now have periods, frequency spreads, and power fractions for each radial order and degree, allowing for direct comparisons between each of the derived modes. To evaluate luminosity variations, we calculate the spherically averaged luminosity at the outer boundary of the computational domain. While spherical averaging may retain some imprints of non-radial modes, these are less distinct and not reliably distinguishable. However, radial orders remain well separated and can be confidently analysed.

\subsection{Radial node extraction process}\label{sect_method_nodes}

   \begin{figure}[t]
   \centering
   \includegraphics[width=\hsize]{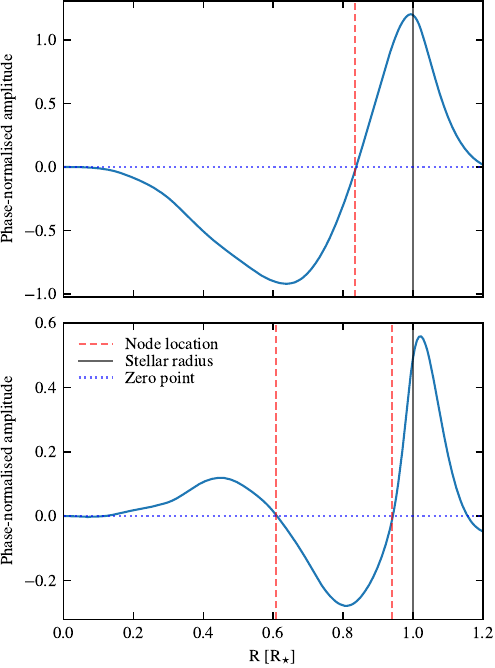}
      \caption{The phase-normalised amplitude derived from the Fourier transform of the  quantity $\sqrt{\rho}v_{\mathrm{radial}}$, in the frequency bands of the first and second overtones (for model \texttt{C}), in the top and bottom row, respectively.
              }
         \label{fig_examplenodes}
   \end{figure}
%

We further utilise the velocity data within the star to identify radial nodes when an overtone is detected using the process described above. First, we take the spherically averaged quantity $\sqrt{\rho}v_{\mathrm{radial}}(r,t)$ (see Sect.~\ref{sect_method_pulsation_extraction}) as an effective measure of density variations within the star. The complex Fourier transform of this quantity is calculated, now as a function of radial distance and frequency, and subsequently used in the following steps.

The node extraction process begins with phase alignment, where the phase of the complex Fourier-transformed signal is shifted such that the phase at the stellar radius is set to zero. This is implemented by dividing each radial Fourier profile by the complex phase at the radius, leaving only the relative phase differences across the radial layers. For each identified radial mode (fundamental or overtone), a frequency mask defined by $f_{\textrm{mean}} \pm f_{\textrm{RMS}}$ was applied to isolate the corresponding spectral window, which was derived in the step outlined in Sect.~\ref{sect_method_pulsation_extraction}. This ensures that only the signal associated with the mode of interest is considered in the subsequent steps.

At each frequency within the masked range, the Fourier-transformed signal provides a radial profile of the oscillation amplitude. These profiles are then averaged across the frequency window to produce a single, representative radial profile for the mode. We refer to this averaged signal as a phase-normalised amplitude. This averaged profile is still complex-valued, and its real part predominantly captures the amplitude of the standing wave of the pulsation mode.
We extract the real part of this profile and interpret the sign to determine the relative phase of the pulsation at each radius: positive values indicate in-phase motion with the stellar surface, while negative values indicate anti-phase behaviour. Radial nodes are then identified as the locations where this profile crosses zero, marking phase transitions. Figure~\ref{fig_examplenodes} shows examples of these phase-aligned amplitude profiles for the first and second overtone modes of model \texttt{C}. The non-zero signal extending beyond the stellar radius reflects travelling waves that turn into outward-propagating shocks, as discussed by \citet{freytagBoundaryConditionsCO5BOLD2017} and \citet{ahmadPropertiesSelfexcitedPulsations2023a}.


   \begin{figure}[t]
   \centering
   \includegraphics[width=\hsize]{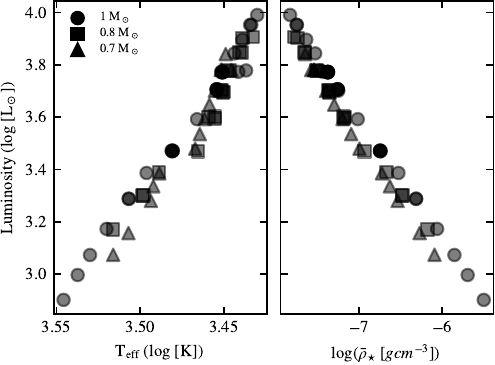}
      \caption{Luminosity plotted against effective temperature and mean density, illustrating the distribution of models within the parameter space.
              }
         \label{fig_paramspace}
   \end{figure}
%

\section{Results}\label{resultssection}

The model grid was designed to expand the parameter space, focusing on higher-density AGB stars, as illustrated in Fig.~\ref{fig_paramspace}. It extends beyond the range of our previous work \citep[in particular, the models considered in][]{freytagGlobal3DRadiationhydrodynamics2017, ahmadPropertiesSelfexcitedPulsations2023a}.

For clarity, the pulsation modes presented and discussed in this work include modes of different radial orders $n=0,1 \, \mathrm{and} \, 2$, corresponding to the fundamental mode, first overtone (1O), and second overtone (2O), respectively\footnote{An electronic table, listing the global stellar parameters of all models together with the derived pulsation periods, is available in electronic form at the Centre de Données astronomiques de Strasbourg (CDS).}. There are also non-radial modes, denoted by their degree ($\ell$). Radial modes are characterised by $\ell=0$, with $\ell=1$ representing dipole modes and $\ell=2$ representing quadrupole pulsation modes.

\begin{figure*}[b]
\resizebox{\hsize}{!}
        {\includegraphics[]{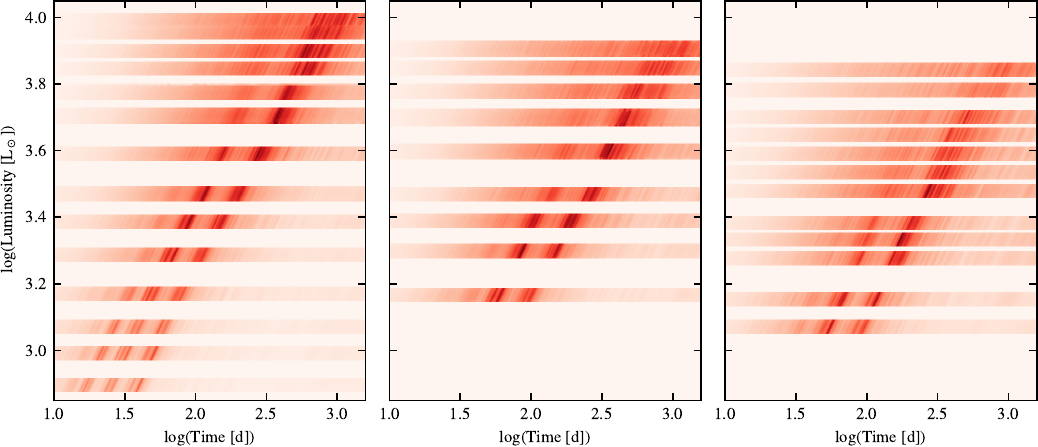}}
  \caption{Period-luminosity plots for the models in our grid, with each panel representing models of different masses: $1\,\mathrm{M}_{\odot}$, $0.8\,\mathrm{M}_{\odot}$, and $0.7\,\mathrm{M}_{\odot}$ from left to right. The colour bar indicates the summed power of the radial, dipole, and quadrupole velocity power spectra at the stellar surface; with darker colours representing higher values. Note that the rows are artificially slanted by a constant factor to improve visual separation and highlight the progression of spectral features.}
     \label{fig_periodlum-ridge-all}
\end{figure*}


\subsection{Period-luminosity relation}

We present colour-map plots of the combined power spectra for radial, dipole, and quadrupole modes on the stellar surface in Fig.~\ref{fig_periodlum-ridge-all}. The distinct signatures of different pulsation modes are clearly visible, showing consistent qualitative patterns across the three stellar masses.
In the high-luminosity region ($\log(\mathrm{L}_\star \, [\mathrm{L}_\odot])\ge 3.8$), the pulsation signal appears broadly spread, corresponding to the radial fundamental mode, as identified in previous studies \citep[e.g.][]{freytagGlobal3DRadiationhydrodynamics2017, ahmadPropertiesSelfexcitedPulsations2023a}. As luminosity decreases to the midrange ($3.3  \le \log(\mathrm{L}_\star \, [\mathrm{L}_\odot]) \le 3.8$), the fundamental mode becomes more distinct with a narrower spread, while signatures of higher overtones start to emerge. This is also the luminosity range where the pulsation amplitude reaches its peak.
Below $\log(\mathrm{L}_\star \, [\mathrm{L}_\odot])\le 3.3$, the overtones become more prominent, clearly separated from the fundamental mode, and being at shorter periods. In the $1\, \mathrm{M}_{\odot}$ sequence, the second overtone is particularly noticeable, showing clear separation from both the fundamental and first overtone modes.

Overall, the fundamental, first, and second radial overtones can be distinguished clearly, with a smaller scatter within each ridge in Fig.~\ref{fig_periodlum-ridge-all}. 
As the pulsation period decreases with increasing radial order, the corresponding ridges in the P-L diagram can be broadly associated with the observed sequences: the fundamental mode ($n=0$) with sequence C, the first overtone ($n=1$) spanning sequences C' and B, and the second overtone ($n=2$) primarily aligning with sequence A \citep[e.g.][]{woodMACHOObservationsLMC1999b, takayamaPulsationModesOGLE2013, woodPulsationModesMasses2015b, trabucchiModellingLongperiodVariables2019}.
For the fundamental mode, our previous models have been shown to be in agreement with the P-L relation derived from observations of Miras pulsating in the fundamental mode, after applying bolometric corrections \citep{ahmadPropertiesSelfexcitedPulsations2023a}. The fundamental pulsation periods in the current grid, for stars with similar stellar parameters between the previous and current grids, remain consistent despite adjustments to the model settings described in Sect.~\ref{sect_method_differencesinmodels}. Consequently, we do not recompare the P-L relation with observations of Miras in this study.

Within each ridge, smaller-scale separations can be seen, which reflect the imprints of modes with different angular degrees; specifically, the radial, dipole, and quadrupole modes. The process of producing Fig.~\ref{fig_periodlum-ridge-all}, which involved applying the FFT to the different velocity as a function of $\ell$ and $m$, enabled the identification of these smaller separations within each radial order. These modes are further analysed in Sect.~\ref{sect_nonradialmodes}.

   \begin{figure}[tp]
   \centering
   \includegraphics[width=\hsize]{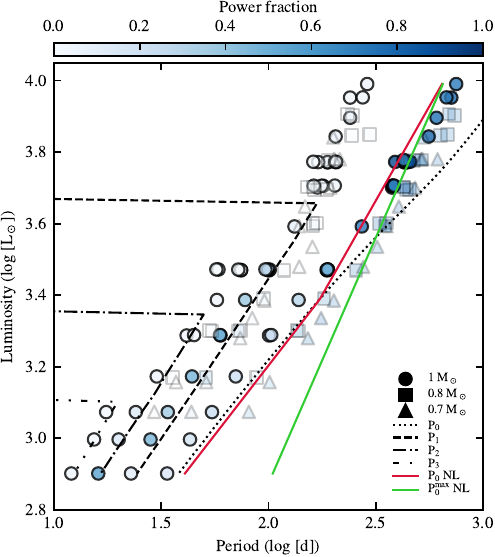}
      \caption{The derived P-L relation, where the colour represents the fraction of power in the power spectra, derived from the FFT of the radial velocity, on the stellar surface for each identified mode in the models. Data points for $0.7 \, \mathrm{and} \, 0.8\, \mathrm{M}_{\odot}$ models are faded to emphasise the $1\, \mathrm{M}_{\odot}$ models. The lines with different patterns correspond to the linear fundamental and first-, second-, and third-overtone periods ($\mathrm{P}_0, \mathrm{P}_1, \mathrm{P}_2 \, \mathrm{and} \, \mathrm{P}_3$, respectively) interpolated from \citet{trabucchiModellingLongperiodVariables2019}. The non-linear fundamental period and its maximum limit (respectively $\mathrm{P}_0 \, \mathrm{NL} \, \mathrm{and} \, \mathrm{P}_0^{max} \, \mathrm{NL}$) follows the analytical formulae derived from \citet{trabucchiModellingLongperiodVariables2021a}.
              }
         \label{fig_periodlum-1dcomparison}
   \end{figure}
%

   \begin{figure}[t]
   \centering
   \includegraphics[width=\hsize]{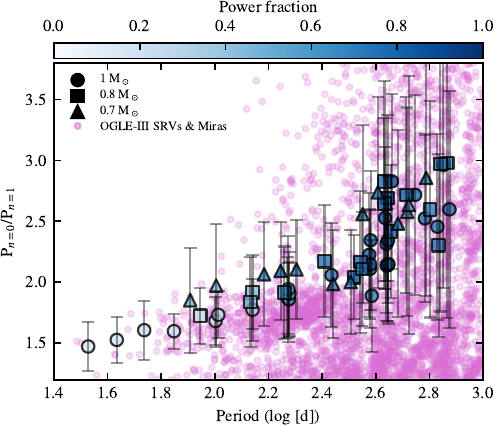}
      \caption{Ratio of the fundamental to first overtone periods plotted against the fundamental mode period. The colour represents the fraction of power of the fundamental mode, in the power spectrum of the radial velocity, on the stellar surface. The smaller circles are stars identified as either SRVs or Miras from the OGLE-III catalogue (see text for details).
              }
         \label{fig_nthradialmode}
   \end{figure}
%

\subsection{Comparison with 1D models}\label{sect_compare1d}

To compare our 3D models with existing 1D models, we utilised the interpolation routine from \citet{trabucchiModellingLongperiodVariables2019} and the non-linear fundamental mode periods from \citet{trabucchiModellingLongperiodVariables2021a}. For the interpolation, we adopted the parameters of our $1 \, \mathrm{M}_{\odot}$ sequence, incorporating its respective stellar parameters (core mass of $0.6 \, \mathrm{M}_{\odot}$, effective temperature, and luminosity). The mass fractions of hydrogen, helium and other metals ($X=0.708768$, $Y=0.275033$, and $Z=0.016199$) were consistent with those used in the \texttt{CO5BOLD} simulations, and a carbon-to-oxygen ratio of 0.48 was set, following \citet{liljegrenPulsationinducedAtmosphericDynamics2017b}.

We compare the P-L relations from our 3D models with the predictions of 1D models, as shown in Fig.~\ref{fig_periodlum-1dcomparison}. Overall, the $1\, \mathrm{M}_{\odot}$ models align well with 1D predictions regarding the positioning of radial order ridges in the P–L plane. However, in the 3D models, higher overtone modes appear at higher luminosities compared to their 1D counterparts, particularly for the first and second overtones. For the fundamental mode, our 3D models support the presence of a "break" in the P–L relation, consistent with the non-linear models of \citet{trabucchiModellingLongperiodVariables2021a}. This result highlights that while linear 1D models effectively predict periods of overtone modes, they systematically overestimate fundamental mode periods in Mira-like stars due to their inability to account for large-amplitude pulsations. Such pulsations significantly alter the stellar structure, causing deviations from hydrostatic equilibrium and necessitating a non-linear treatment.

The differences between the fundamental mode periods derived from our 3D models and those predicted by 1D models, as shown in Fig.~\ref{fig_periodlum-1dcomparison}, are most pronounced at both the high- and low-luminosity ends of the model sequence. At high luminosities, the 3D periods tend to be longer than their 1D counterparts, likely reflecting non-linear effects. Conversely, in the low-luminosity regime, the 3D periods are systematically shorter than those predicted by 1D models. The discrepancy likely stems from structural differences between the 1D and 3D models (e.g. the stellar radii that each predicts). Further elucidating these differences would require more detailed comparisons and parameterisation efforts.

In terms of amplitudes, our most Mira-like model has a bolometric amplitude of $\Delta m_\mathrm{bol} \approx 0.2\,\mathrm{mag}$, significantly lower than the typical amplitudes of Mira stars observed by \citet{whitelockInfraredColoursMiralike2000}, which exceed $\Delta m_\mathrm{bol} \approx 0.5 \,\mathrm{mag}$. In contrast, the amplitudes predicted by \citet{trabucchiModellingLongperiodVariables2021a} tend to be higher, with the amplitudes being sensitive to the turbulent viscosity parameter. These differences highlight the challenges in reproducing observed amplitudes across different modelling approaches. The interplay between convection and pulsation is evident in the pulsation amplitudes of our AGB models. This interaction is further explored in Sect.~\ref{sect_persistenceofmultimodes}.

\subsection{Radial modes and nodes}\label{sect_radialnodes}

\begin{figure*}[t!]
\resizebox{\hsize}{!}
        {\includegraphics[]{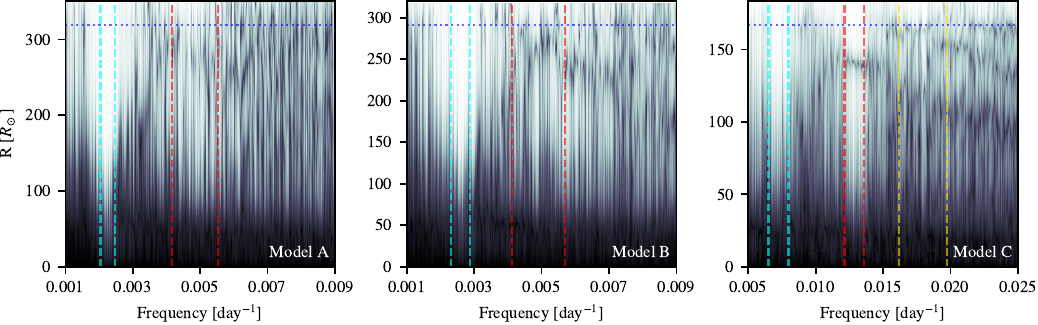}}
  \caption{The power spectrum as a function of radial distance and frequency for representative models \texttt{A}, \texttt{B}, and \texttt{C} (see text for details). Histogram equalisation is applied to each power spectrum to enhance signal contrast. In each panel, the enclosed vertical lines indicate the frequency band (${f_{\mathrm{mean}}} \pm f_{\mathrm{RMS}}$) of the fundamental mode and higher overtones, ordered from low to high frequency. The horizontal dotted-line marks the stellar radius of the respective model. }
     \label{fig_3dfftspectra}
\end{figure*}

   \begin{figure}[tp]
   \centering
   \includegraphics[width=\hsize]{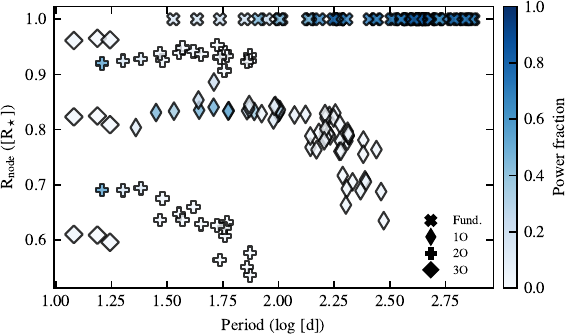}
      \caption{The location of nodes, expressed as a fraction of the stellar radius, is shown for all the models presented in this work. Symbols indicate the node corresponding to the radial order: the first overtone has a single internal node, the second overtone has two nodes, and the third overtone has three nodes. The fundamental mode, which has no internal nodes, is included to represent its node at the stellar surface. The colour represents the fraction of power in the power spectrum, derived from the FFT of the radial velocity at the stellar surface, for each identified mode in the models.
              }
         \label{fig_nthradialnodelocations}
   \end{figure}
%

We examined the period ratio between the fundamental and first overtone modes, in the radial mode, as shown in Fig.~\ref{fig_nthradialmode}. For our models, we derive that where the ratio lies between $1.7 - 2.0$, the models in this range corresponds to models with stellar luminosities of approximately between $3.4 \le \log(\mathrm{L}_\star \, [\mathrm{L}_\odot]) \le3.6$. Above this luminosity range, the period ratios increase significantly. Although overtone modes may still be detectable in this regime, their reliability diminishes, as indicated by their minimal power fraction in the FFT power spectrum, where the fundamental mode dominates. Furthermore, the fundamental mode exhibits a relatively larger spread in its signal. This, combined with weaker and more dispersed signals from overtone modes, results in greater uncertainty in the calculated period ratios.

The spreads in both the fundamental and first overtone modes further affect the relative uncertainties relevant for this analysis, but most of the calculated ratios intercept the range of $1.7 - 2.0$. This range of ratio is interesting, as the models in this range aligns with the expected P-L relation between the fundamental and overtone modes in AGB stars. The large period ratios between the fundamental and first overtone modes in AGB stars are a natural consequence of their internal structure. This structure leads to large period ratios, typically in the range of $1.7 - 2.0$ \citep[e.g.][]{beddingModeSwitchingNearby1998, takeutiMethodEstimateMasses2013a, fuentes-moralesMultiperiodicSemiregularVariable2014}. Stars pulsating in the fundamental mode generally have much smaller radii than those of the same mass and luminosity pulsating in an overtone. Observationally, stars such as R Doradus, with a period ratio of 1.81, exemplify this behaviour \citep{beddingModeSwitchingNearby1998}. This narrow range of period ratios is consistent with the conclusion that the stars are pulsating in the same pair of modes.

Additionally, in Fig.\ref{fig_nthradialmode}, we include SRVs and Miras from the OGLE-III catalogue of LPVs in the Large Magellanic Cloud (LMC) \citep{2009AcA....59..239S}. The aim is to visualise the observed spread in the ratio of the fundamental to first overtone period, and to compare this with our models. To this end, we derived a subset of 3429 sources (3201 SRVs and 228 Miras) by restricting the OGLE-III dataset to those that fall within the axis limits of Fig.~\ref{fig_nthradialmode}. While the OGLE-III catalogue provides amplitude-based mode assignments \citep[as analysed in][]{woodPulsationModesMasses2015b}, we instead order the periods for each star by length, assuming the longest period corresponds to the fundamental mode, followed by the first overtone.
This comparison highlights that the spread in the period ratios produced by our models aligns well with the observed distribution of LPVs, lending further confidence to the physical realism of the simulations. We note, however, that the assumption of mode ordering by period introduces some uncertainty: in particular, the observed spread may partially reflect differences in angular degree, such as the fundamental mode being radial while the overtone corresponds to a non-radial mode (e.g. dipole). Nonetheless, the similarity in the spread of the period ratios reinforces the consistency between the models and observations.

To highlight the presence of radial nodes in our models and complementary to Fig.~\ref{fig_examplenodes}, Fig.~\ref{fig_3dfftspectra} presents the power spectra of three representative models, all corresponding to $1\,\mathrm{M}_{\odot}$ stars exhibiting distinct self-excited pulsation behaviours as described in Sect.~\ref{sect_method_differencesinmodels}. The properties of their simulated luminosities are examined in Sect.~\ref{section_results_wavelets}. In the power spectra presented in Fig.~\ref{fig_3dfftspectra}, nodes appear as minima in power, with the node corresponding to the first overtone in model \texttt{C} being particularly prominent. 

The distribution of radial nodes in our models is shown in Fig.~\ref{fig_nthradialnodelocations}, where their locations align as expected with pressure-driven oscillations (p-modes) in spherically symmetric systems. As the radial order $n$ increases by one, a new radial node appears, resulting in a higher pulsation frequency of the corresponding overtone. The radial eigenfunctions, describing the displacement of material in the radial direction, exhibit standing wave patterns within the star. With increasing radial order, the innermost node shifts closer to the stellar centre, while the outermost node extends further toward the stellar surface, reflecting the increased spatial segmentation required to maintain orthogonality between modes. A notable result from Fig.~\ref{fig_nthradialnodelocations} is that the node of the first overtone and the lower node of the second overtone extend deeper for longer pulsation periods than those of shorter periods in the same mode. These longer pulsation periods correspond to higher-luminosity, lower-mean density stars. This may reflect the strength of convection and how deeply the downdrafts penetrate into the stellar interior, a phenomenon discussed further in Sect.~\ref{sect_persistenceofmultimodes}.

\subsection{Mean density relation}\label{sect_meandensityrelation}

The relationship between the pulsation period and the mean density of a star has long been a cornerstone of stellar pulsation theory. \citet{ritterUntersuchungenUeberHoehe1879} first established this relationship for a pulsating homogeneous sphere undergoing adiabatic radial pulsations. The resulting relation, $P_{\textrm{puls}}\propto \bar{\rho}_{\star} ^{-0.5}$, links the pulsation period to the inverse square root of the mean density of the sphere. Applied to stars, this implies that their contraction and expansion are accompanied by periodic variations in effective temperature and luminosity. In \cite{ahmadPropertiesSelfexcitedPulsations2023a}, we demonstrated that our models do follow the mean density relation well, an intriguing result given that the relation is derived under the assumptions of a homogeneous and spherically symmetric atmosphere undergoing adiabatic pulsations. We also showed that the pulsation period of a radial mode can be estimated by the sound-crossing time, defined as the time required for a sound wave to propagate from the centre to the surface of the star and back.


   \begin{figure}[t]
   \centering
   \includegraphics[width=\hsize]{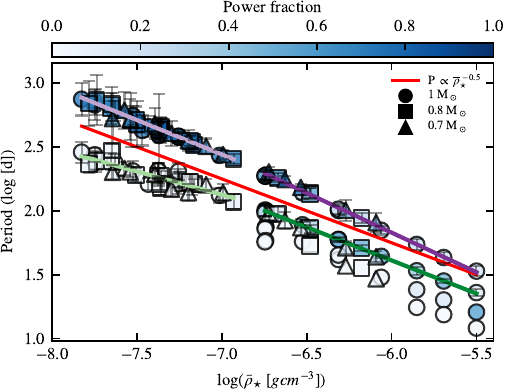}
      \caption{Derived periods of radial modes plotted against the stellar mean density. The red line represents Ritter's period-mean density relation, while the other coloured lines are fitted relations for radial orders $n=0 \, \mathrm{and} \, 1$, with their results detailed in Table~\ref{table:gradient_y_intercept}. The colour represents the fraction of power in the power spectra, derived from the FFT of the radial velocity, on the stellar surface for each identified mode in the models.
              }
         \label{fig_meandensityrelation}
   \end{figure}
%

\begin{table}[t]
\caption{Resulting linear regression parameters for the period-mean density ridges.}  
\label{table:gradient_y_intercept}                          
\centering                                                  
\begin{tabular}{c c c c}                                    
\hline\hline                                                
n & Segment & Gradient & Intercept  \\                     
\hline                                                     
0 & Whole  & $-0.592 \pm 0.012$ & $-1.712 \pm 0.082$ \\ 
0 & Left   & $-0.539 \pm 0.052$ & $-1.328 \pm 0.375$ \\ 
0 & Right  & $-0.621 \pm 0.023$ & $-1.898 \pm 0.149$ \\ 
1 & Whole  & $-0.474 \pm 0.008$ & $-1.228 \pm 0.049$ \\ 
1 & Left   & $-0.364 \pm 0.044$ & $-0.421 \pm 0.325$ \\ 
1 & Right  & $-0.520 \pm 0.017$ & $-1.508 \pm 0.105$ \\ 
\hline                                                     
\end{tabular}
\tablefoot{Gradients and intercepts obtained from linear regressions applied to successive segments along the fundamental- and first-overtone ridges in the period-mean density plane (see Fig.~\ref{fig_meandensityrelation}).}
\end{table}

Our models presented here generally conform well to the period–mean density relation, as demonstrated in Fig.~\ref{fig_meandensityrelation}. A notable result is that the agreement with this relation depends on which mode is dominant. Specifically, the fundamental mode dominates at lower mean densities, while the first overtone becomes dominant at higher mean densities. These differences are indicated by the filled colours of the points in Fig.~\ref{fig_meandensityrelation} and are more clearly distinguished in the top panel of Fig.~\ref{fig_pulsmodedom}.

The shift in mode dominance is reflected in the gradients of the period–mean density plane, as quantified in Table~\ref{table:gradient_y_intercept}. For subsets of models below and above a mean density of $\log(\bar{\rho}_\star \, [\mathrm{gcm}^{-3}]) = -6.8$, the gradient varies depending on which mode dominates. This density threshold corresponds to an average luminosity of $\log(\mathrm{L}_\star \, [\mathrm{L}_\odot]) \approx 3.5$ in our models, marking a transition in the pulsation behaviour. At lower mean densities, the fundamental mode provides a better match to the period–mean density relation, whereas at higher densities, the first overtone aligns more closely. These results suggest a mode-dependent scaling of periods with mean density that reflects changes in the star's internal structure.

\subsection{Comparing radial and non-radial modes}\label{sect_nonradialmodes}

   \begin{figure}[t]
   \centering
   \includegraphics[width=\hsize]{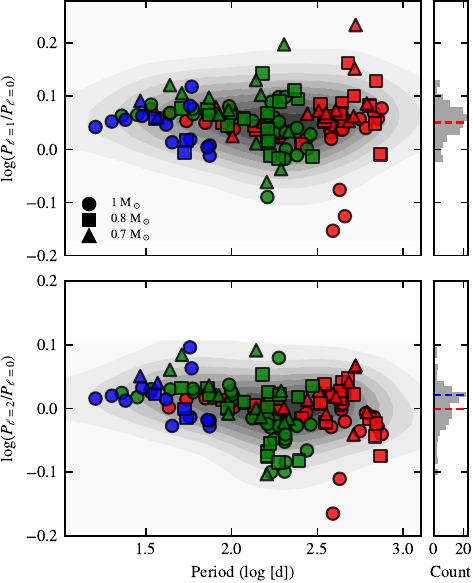}
      \caption{Comparison of derived pulsation periods for different degrees. Top panel: Logarithmic ratio of periods derived from $\ell=2 \, \mathrm{and} \, \ell=0$. Bottom panel: Logarithmic ratio of periods derived from $\ell=1 \, \mathrm{and} \, \ell=0$. The colours indicate pulsation modes in the fundamental, first and second overtones, represented by red, green, and blue, respectively. The contours in each main panel illustrates the data point distribution. The subpanels in each row show histograms of the ratios, with dashed lines marking the mean values. For the bottom panel, the blue dashed line in the histogram represents the mean ratio when a luminosity criterion of $\log(\mathrm{L}_\star \, [\mathrm{L}_\odot])\le 3.5$ is applied to the models being considered for the corresponding ratio results.
              }
         \label{fig_pulsperratio}
   \end{figure}
%

\subsubsection{Period hierarchy of modes}

Previous studies on non-radial pulsations of red giants have established characteristic period relationships among pulsation modes. These works \citep[e.g.][]{dziembowskiOscillationsUMaOther2001, mosserPeriodluminosityRelationsEvolved2013, woodPulsationModesMasses2015b, montalbanNonradialModesAGB2017} consistently show that for a given radial order, the dipole mode generally has the longest period, the radial mode has the shortest period, and the quadrupole mode has a period which falls in between. To test this in our models, we compare the periods of dipole and quadrupole modes to the radial mode for radial orders up to $n=2$, as presented in Fig.~\ref{fig_pulsperratio}.

\begin{figure*}[tp]
\resizebox{\hsize}{!}
        {\includegraphics[]{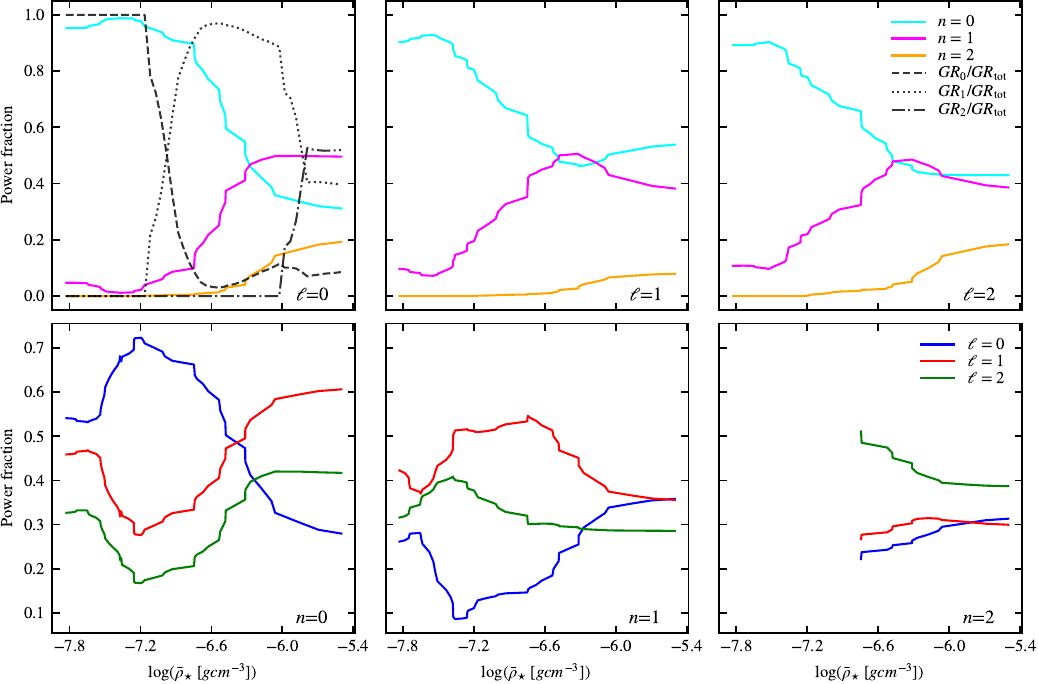}}
  \caption{Relative power fractions of the radial, dipole, and quadrupole modes as a function of stellar mean density. The top row shows the relative power distribution across radial orders for each angular degree $\ell$, calculated as ${P^\ell _n} / {\sum _{\ell'} P ^{\ell'}_n}$. The black-lines in the leftmost panel indicate the fractions of the growth rates of the fundamental, first and second overtones relative to summed-up growth rates, providing a means of comparing the relative amplitudes derived from linear 1D models \citep{trabucchiModellingLongperiodVariables2019}, as described in Sect.~\ref{sect_compare1d}. The bottom row displays the relative power for each angular degree at a given radial order $n$, calculated as ${P^\ell _n}/{\sum {n'} P ^{\ell}{n'}}$. These plots illustrate how the dominant mode shifts depending on the mean density and radial order. Note that all power values are evaluated at the stellar surface for each identified mode.}
     \label{fig_pulsmodedom}
\end{figure*}

Our results confirm that the dipole mode consistently has the longest period, with an average value approximately $12 \, \%$ greater than the radial mode. For the quadrupole mode, when considering all models, the period appears close to that of the radial mode on average. However, when restricting the analysis to models with $\log(\mathrm{L}_\star \, [\mathrm{L}_\odot]) \le 3.5$, the quadrupole mode exhibits an average period approximately $6\, \%$ higher than the radial mode. At higher luminosities (i.e. lower densities), discrepancies arise in the period hierarchy. The reduced surface gravity and extended stellar atmosphere lead to greater dominance of the radial mode and larger spreads in the pulsation modes. This behaviour likely makes the periods derived from the $\ell=2$ velocity less distinguishable, bringing them closer to those of the radial mode $\ell=0$ and even possibly blending the period into the radial mode. Given that the pulsation modes generally spread out more going towards higher luminosities, it offers an explanation on why we did not observe strong or identifiable imprints other than the radial modes in our previous model grids \citep[e.g. in][]{freytagGlobal3DRadiationhydrodynamics2017, ahmadPropertiesSelfexcitedPulsations2023a}.
Given this consideration, restricting the analysis of the quadrupole-to-radial period ratios to models with $\log(\mathrm{L}_\star \, [\mathrm{L}_\odot]) \le 3.5$ provides a more sensible analysis. With this restriction applied and disregarding higher-degree modes ($\ell = 3, 4, 5, \, \ldots$), our results align with the established trend, being that the dipole mode has the longest period, the radial mode has the shortest, and the quadrupole mode falls in between.

\subsubsection{Dominant pulsation modes}

We investigate the dominant pulsation modes across the model grid, as shown in Fig.~\ref{fig_pulsmodedom}. We calculate the relative fraction of power associated with each mode: the radial, dipole, or quadrupole modes, and across the radial orders. In the top row, the fraction is calculated as ${P^\ell _n} / {\sum _{\ell'} P ^{\ell'}_n}$, representing the relative power distribution across radial orders for each angular degree $\ell$. In the bottom row, the fraction is calculated as ${P^\ell _n}/{\sum _{n'} P ^{\ell}_{n'}}$, showing the relative power for each angular degree at a given radial order $n$. These relative power fractions are plotted as a function of the stellar mean density.

In the top row of Fig.~\ref{fig_pulsmodedom}, the strengthening of overtone modes with increasing mean density across different angular degrees is clearly illustrated. To compare these findings with the 1D results \citep[derived from results of][]{trabucchiModellingLongperiodVariables2019, trabucchiModellingLongperiodVariables2021a} presented in Sect.~\ref{sect_compare1d}, we include the ratios of growth rates as derived in the linear models, in the first column of the first row, as the dominant mode is typically associated with the largest growth rate. This comparison reveals strong agreement with our 3D results, particularly in capturing the transition where the fundamental mode loses dominance and the first overtone mode gains strength as a function of mean density, eventually becoming the dominant pulsation mode. Similar trends are observed when the fractions are evaluated based on the spherically averaged luminosity at the edge of the computational domain. Despite lacking information about the angular degree of the pulsations, this approach still captures the fact that the first overtone mode is stronger at higher mean densities.

The bottom row of Fig.~\ref{fig_pulsmodedom} helps identify which degree dominates within each radial order. At lower mean densities, the radial mode is dominant, while the dipole mode becomes dominant at higher mean densities. This shift in dominance is more pronounced for the first overtone mode, where the dipole modes shows the largest amplitudes across the entire range of mean densities. For the second overtone, the quadrupole mode dominates, although the parameter space is limited. The lower-mean density models do not pulsate strongly in the second overtone, and may be deemed as not present in the method to extract the pulsations. This limits the analysis and the ability to consider them as reliable.

\begin{figure*}[bp]
\resizebox{\hsize}{!}
        {\includegraphics[]{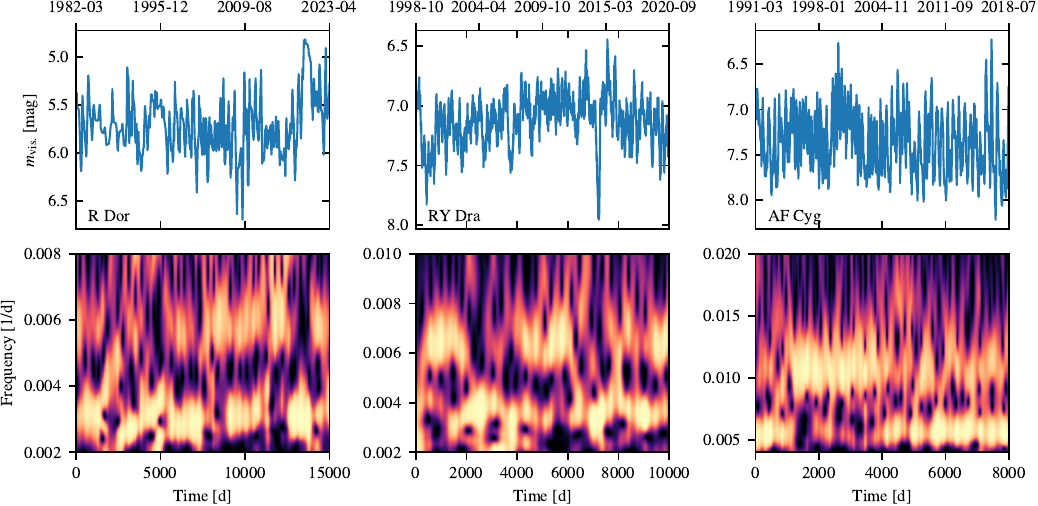}}
  \caption{Visual magnitude observations from the AAVSO database (top row) and the corresponding WWZ transform applied to the observations (bottom row). The WWZ transform is vertically normalised, such that each column in the plot sums to 1. From left to right, the columns represent R Dor, RY Dra, and AF Cyg, respectively. 
          }
     \label{fig_wavelet-obs}
\end{figure*}

\subsection{Wavelets comparison between observations and models}\label{section_results_wavelets}

We use the weighted wavelet Z-transform (WWZ) method \citep{fosterWaveletsPeriodAnalysis1996}, as adapted by \citet{kiehlmannWwzWeightedWavelet2023}, to study the complex behaviour of multi-mode pulsations. The WWZ analysis is particularly effective for examining phenomena such as frequency modulation, amplitude modulation, phase jumps, and mode-switching \citep[e.g.][]{szatmaryApplicationWaveletAnalysis1994, templetonSecularEvolutionMira2005}. Its ability to handle irregularly sampled data makes it ideal for observations in datasets like the AAVSO, which often lack consistent temporal coverage.

We applied the WWZ transform to identify mode-switching and overtone activity in pulsations, focusing on three well-studied stars: R Dor, RY Dra, and AF Cyg. These stars were selected for their distinct pulsation behaviours. R Dor, a Mira variable, exhibits overtone activity \citep[see][]{beddingModeSwitchingNearby1998}, while RY Dra and AF Cyg are semiregular variables of type SRb, oxygen-rich and carbon-rich, respectively. All three stars show mode-switching behaviour and irregular pulsations \citep[e.g.][]{kissMultiperiodicitySemiregularVariables1999b, hinkleVelocityObservationsMultipleMode2002, percyAmplitudeVariationsPulsating2013a}.

\begin{figure*}[t]
\resizebox{\hsize}{!}
        {\includegraphics[]{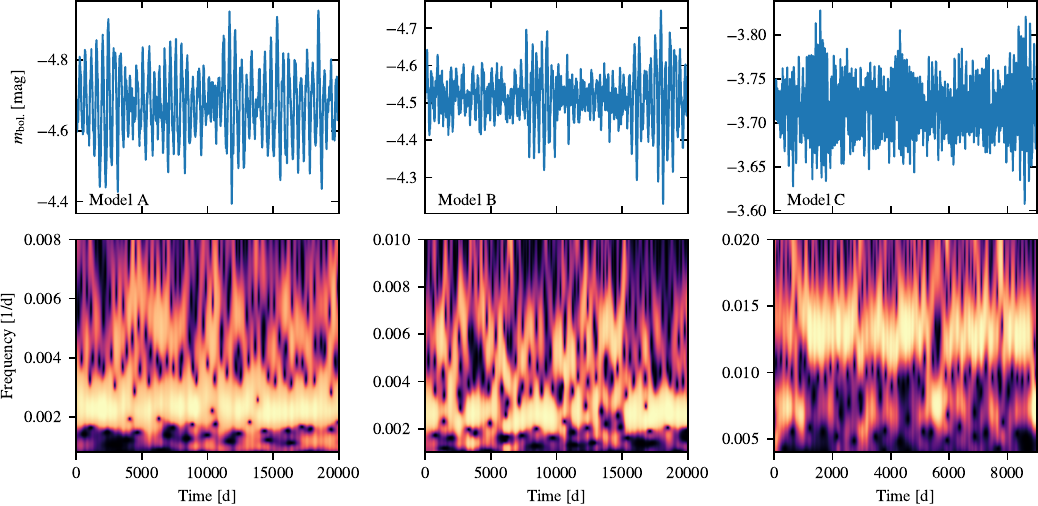}}
  \caption{Bolometric magnitudes (transformed from luminosity) plotted across simulation time (top row) and the corresponding WWZ transform (bottom row). The WWZ transform is vertically normalised, ensuring each column in the plot sums to 1. From left to right, the columns represent models A, B, and C, respectively (see text for details).}
     \label{fig_wavelet-models}
\end{figure*}

Our analysis here focuses on mode-switching behaviour and pulsation periods, as a detailed investigation of pulsation modulation and power is beyond the scope of this work. The WWZ transform was applied to the visual light curves of R Dor, RY Dra, and AF Cyg, as acquired from the AAVSO database, with results shown in Fig.~\ref{fig_wavelet-obs}. To compare with the observations, we applied the WWZ transform to the time-series luminosity of the three representative models described in Sect.~\ref{sect_method_differencesinmodels}. Their time-series luminosity and respective WWZ spectra are displayed in Fig.~\ref{fig_wavelet-models}.

The WWZ spectrum for R Dor (Fig.~\ref{fig_wavelet-obs}, left column) reveals strong variability in the first overtone mode, while the fundamental mode remains consistent over the chosen timeframe. Model \texttt{A} (Fig.~\ref{fig_wavelet-models}, left column) exhibits a similar pattern, with the fundamental mode dominating but accompanied by weaker, noticeable overtone signals. The fundamental and overtone frequencies in the model align well with those observed in R Dor.

For RY Dra, the WWZ spectrum indicates active mode-switching, with irregular transitions between the fundamental and first overtone modes. Model \texttt{B} displays comparable behaviour in its WWZ spectrum, with a persistent but less consistent fundamental mode relative to model \texttt{A}. Between 10,000 and 15,000 days of simulated time, the fundamental mode in model \texttt{B} dampens, while the first overtone gains dominance, a pattern that mirrors the observed behaviour of RY Dra.

AF Cyg, known for its low-amplitude pulsations, exhibits strong and persistent first overtone activity alongside weaker fundamental mode pulsations. This contrasts with R Dor, where the fundamental mode dominates. Model \texttt{C} demonstrates a clear dominance of the first overtone mode, with weaker and sporadic signals from the fundamental mode. The pulsation frequencies in Model \texttt{C} are in general agreement with those observed in AF Cyg.

The WWZ analysis effectively identifies mode-switching and overtone activity in both observational data and models. The qualitative agreement between observed stars and the self-excited behaviours in the models highlights the potential of this method for studying the complex dynamics of pulsating stars. However, discrepancies in mode dominance and persistence, especially in stars with irregular pulsations, suggest areas for further investigation. These discrepancies may stem from the high uncertainties of the inferred stellar parameters of the observed stars, as well as differences from the model parameters. Factors such as core mass uncertainty, the absence of radiative pressure in the core (which may become significant with a higher-mass core or increased core temperature during the contraction phase of a pulsation), and certain model assumptions (detailed in Sect.~\ref{sect_method_3dmodels}) may all contribute to these deviations.

\section{Discussion}

\subsection{Mode dominance and evolutionary trends}

The pulsation modes of AGB stars are intrinsically tied to their evolutionary stages \citep[see][]{riebelInfraredPeriodLuminosityRelations2010, takeutiMethodEstimateMasses2013a, mcdonaldOnsetAGBWind2019}. As stars progress along the AGB, changes in luminosity, radius, and mean density associated with their evolution lead to shifts in mode dominance. The fundamental mode tends to dominate in more luminous stars with larger radii, while overtone modes are favoured in denser, less luminous stars during the earlier phases of AGB evolution.

This relationship is evident in our models and aligns well with observed P-L sequences. The top panel of Fig.~\ref{fig_pulsmodedom} illustrates the stronger prominence of the first overtone mode at higher stellar mean density. As stars expand along the AGB and their mean density decreases, the dynamic interplay between pulsation modes becomes more pronounced. In transitional regions, where the first overtone loses strength while the fundamental mode gains amplitude, mode switching is likely to occur. This behaviour reflects the complex evolution of pulsation dynamics in response to changes in stellar structure during this phase, with individual stars evolving across different P–L sequences.

\subsection{Mode switching between radial and non-radial modes}\label{sect_discuss_nonradialmodesinteractions}

Mode-switching behaviour is not confined to radial modes; it can also involve transitions between radial and non-radial modes or among different non-radial modes. As AGB stars evolve, changes in surface gravity and density can influence which modes are favoured. For instance, stars with lower surface gravity and extended envelopes may damp non-radial modes, while denser stars are more likely to exhibit higher-degree modes, particularly the dipole mode. The presence of these pulsation modes can lead to constructive or destructive interference, resulting in variability in pulsation amplitudes and affecting light curves and spectral features. These mechanisms are further discussed in Sect.~\ref{sect_persistenceofmultimodes}. Such amplitude modulation and variability are characteristic of semiregular variables, with such variability possibly driven by interactions between convection, radial and non-radial modes \citep{kissMultiperiodicitySemiregularVariables2000a}.

As shown in Fig.~\ref{fig_pulsmodedom}, our grid of models reveals systematic trends in the dominant pulsation modes as a function of stellar parameters. 
In our framework \citep[which follows definition of the sequences described in][]{trabucchiNewInterpretationPeriod2017}, the fundamental mode corresponds to sequence C, the first overtone spans sequences C' and B, and the second overtone is associated with sequence A.
Observational studies have reported differing results regarding which mode dominates at various luminosities and radial orders. For example, \citet{woodPulsationModesMasses2015b} and \citet{stelloNonradialOscillationsMgiant2014a} find that dipole ($\ell = 1$) modes dominate sequence A, particularly at higher luminosities. This broadly aligns with \citet{yuAsteroseismologyLuminousRed2020}, who also report dipole dominance in overtone sequences. In contrast, \citet{mosserPeriodluminosityRelationsEvolved2013} observe that radial ($\ell = 0$) modes dominate at low pulsation frequencies ($\leq 1.0 \, \mu\mathrm{Hz}$), corresponding to high-luminosity stars, particularly in the fundamental and first overtone sequences.

The discrepancies among these findings are especially notable for the first overtone sequence, where the dominant mode appears to shift as a function of luminosity. Additionally, quadrupole modes often follow radial mode trends due to their closer frequency proximity to each other, therefore may carry a higher uncertainty. Further comparisons with OGLE datasets and other observational catalogues will be essential to clarify these trends and resolve the discrepancies.

Our results, presented in Sect.~\ref{sect_nonradialmodes}, are consistent with observational studies while also highlighting the complexities of mode-switching behaviour in evolved stars. This is a particularly noteworthy outcome, especially given that no parameterisation of pulsations was applied in the 3D models. We find that, in the fundamental mode, the radial mode dominates at lower mean densities (corresponding to higher luminosities), whereas the dipole mode becomes dominant at higher mean densities (i.e. lower luminosities). This trend can be interpreted as the suppression of radial modes in denser models, where more stochastic behaviour results in increased amplitudes of non-radial modes. For both the first and second overtones, however, the dipole mode consistently shows the largest amplitudes across the full range of mean densities (if neglecting the quadrupole mode in the second overtone). This finding aligns with recent observational reports suggesting that non-radial modes, particularly the dipole mode, can dominate in overtone sequences.

In contrast to the velocity-based analysis, when the power spectrum of luminosity is used to evaluate the power ratio, the radial mode consistently appears dominant across the radial orders. This discrepancy can be attributed to the bias introduced by calculating luminosity as a spherically averaged quantity outside the star, which neglects non-radial asymmetries and thereby favours the radial mode. When the power spectrum is computed from the intensity flux, which is emergent at each of the six sides of the computational box, and then averaged over all six sides, the dipole mode emerges as dominant in the overtone modes. This result aligns more closely with those obtained from velocity-based power spectra.

It is worth noting that, in principle, higher-degree non-radial modes ($\ell > 2$) could be included in our analysis, as the application of weights to the spherically averaged velocity allows for this. However, these higher-degree modes typically exhibit lower amplitudes and power, leading to greater uncertainty in their identification. Additionally, they are less accessible observationally in red giants. Given these limitations, we restrict our analysis to non-radial modes up to the quadrupole mode in this work, which we find to be a justifiable approach.

\subsection{Amplitude and excitation of multi-mode pulsations}\label{sect_persistenceofmultimodes}

\subsubsection{Convection and its pressure feedback}\label{sss_convectivegeometry}

\begin{figure}[t]
   \centering
   \includegraphics[width=\hsize]{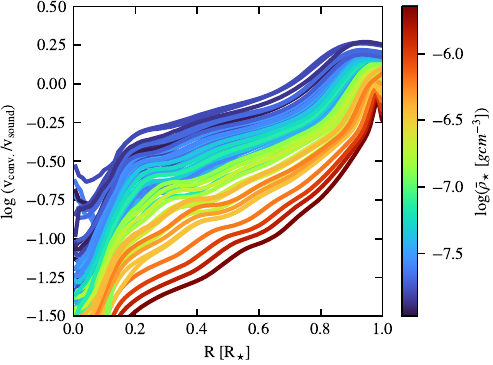}
      \caption{Logarithmic ratio of the convective velocities (computed as the RMS of radial velocities) to the local sound velocity, plotted as a function of the radius (expressed as a fraction of the stellar radius). The lines are coloured according to the corresponding mean density.}
         \label{fig_velocityratio}
   \end{figure}

The surface of the convection-zone of AGB stars is covered by small-scale convection cells analogous to solar granules.  
However, in AGB stars, these cells are often hidden beneath cool, dense clouds of matter just above \citep{freytagDimmingEventsEvolved2024}.  
The downdrafts that originate along the granule borders merge below the surface and form tree-like structures that extend still further inward. Some downdrafts descend all the way from the surface to the model core.  
In this core region, low-order non-radial, long-lived (several pulsation periods) convective flows exist and connect with the downdrafts from above; the latter experience extra acceleration near the core because of the falling gravitational potential. 
The peak velocity in the downdrafts therefore exceeds both the typical up-flow velocity \citep[see Fig.~3 in][]{ahmadPropertiesSelfexcitedPulsations2023a} and the global RMS velocity. 
Figure \ref{fig_velocityratio} plots the ratio of the RMS convective speed to the local sound speed as a function of radius; if the same plot were computed with downdraft velocities, the entire curve would be shifted upward, reflecting the systematically higher downdraft velocities.
Velocities and Mach numbers of convective flows and radial fundamental acoustic
pulsations are large, and their temporal and spatial scales are similar.
Thus, the conditions for interactions between convection and pulsations are favourable \citep{ahmadPropertiesSelfexcitedPulsations2023a}.

In a simple piston model, where the excitation of acoustic pulsations is achieved via the
$\kappa$-mechanism, the opacity is impacted by the periodically-varying temperature and
pressure. This modulates the radiative energy transport leading to temperature
and pressure variations of the gas below the piston in excess to what is
expected from adiabatic pulsations alone.
In our 3D AGB simulations, radiative transport inside the star is negligible, so any modulation predicted by considering $\kappa$-mechanism is inconsequential. Other mechanisms must therefore alter the pressure, evidenced by non-linear changes in pulsation amplitudes.

%

A central pressure enhancement exactly in phase with the adiabatic pressure variation increases the restoring force and accelerates the pulsation (i.e. a faster pulsation).
A pressure increase slightly out of phase could either lead to a driving of the pulsations (if occurring in the expansion phase)
or a damping (if occurring during the compression phase) of the mode.
Convective fluxes vary during a pulsation cycle.
During the compression phase, radial mass, momentum, and energy fluxes are
strongly enlarged.
They all affect the total central pressure.
An enhanced mass flux could increase the gas pressure.
The same happens when the (downward directed) flux of kinetic energy grows.
In contrast, an enhanced (upward) enthalpy flux would decrease the central gas pressure.
Additionally, transport of momentum could increase the central dynamical pressure.

Although the contributions of pressure enhancements and convective fluxes vary roughly in phase with the pulsation, their exact timings differ, depending on the detailed convective geometry.
Short-term changes in the configuration of the large-scale convections cells
can cause strong variations in the pulsations amplitude.
On the other hand, a long-lasting convective dipole flow, for example,
leads to stable, large-amplitude pulsations.

\subsubsection{Mode switching and quantitative diagnostics}
Wavelet analyses of both our models and observations (see Sect.~\ref{section_results_wavelets}) reveal mode-switching behaviour, with overtone modes appearing less stable than the fundamental mode. As stars evolve up the AGB, the interaction of convection and pulsations changes significantly. Pulsations of higher-density, lower-luminosity AGB stars share similarities with solar-like oscillations, where turbulent pressure fluctuations in the outer convective envelope excite the pulsations. The situation differs in more luminous AGB stars. In these stars, deeper and larger-scale convective structures within a lower-density interior become prominent. Strong convective downdrafts can drive high-amplitude radial fundamental modes when in phase with pulsations, but can also suppress overtone modes due to their disruptive nature (as discussed in Sect.~\ref{sss_convectivegeometry}). Conversely, in stars with weaker convection, where motions do not reach as deeply, pressure maxima between the core and surface may help sustain overtone modes by favouring stable oscillatory patterns.

To elucidate the differences in the timescales of the convective and acoustic pulsations, Fig.~\ref{fig_velocityratio} shows the logarithmic ratio of convective velocity (computed as the RMS of radial velocities) to the local sound speed as a function of radius. 
In the dense interior convective speeds are low and sound speeds high, whereas near the surface convective speeds rise and sound speeds fall.  
Stars of higher mean density therefore display lower convective velocities and higher sound speeds throughout.  
Altogether, Fig.~\ref{fig_velocityratio} suggests that overtone amplitudes persist where convective speeds remain well below the local sound speed; however, it is difficult to determine a precise threshold for this.

Surface effects further influence pulsation behaviour. Differences in sound speed between convective regions and adjacent layers can cause scattering of sound waves, which distorts the waves and reduces their coherence. This scattering effect would be pronounced for overtone modes, given their higher frequencies, than for the fundamental mode, making overtone modes particularly vulnerable to disruption.

\subsubsection{Sustaining radial nodes and an analogy with Cepheids}

The amplitude of multiple pulsation modes depends on a balance between excitation and damping processes. 
\cite{macleodLeftRingingBetelgeuse2023a} suggested that the persistence of overtone modes in Betelgeuse depends on the survival of radial nodes, and that strong-enough convection may erase these nodes and trigger mode switching.
By extending this idea, the high convective velocities sustained in AGB stars may destabilise any radial nodes; contributing to mode-switching and irregular pulsation behaviour. This in part may help explain the observed mode-switching in evolved red giants.

A useful comparison can be made with Cepheids, where convection is confined
to relatively shallow layers below the stellar surface, with this depth increasing for lower effective temperatures. In cooler Cepheids, higher convective efficiency dampens overtone modes, while in hotter Cepheids, reduced convective velocities allow overtone modes to dominate \citep{bonoClassicalCepheidPulsation1999}. Although the pulsation excitations mechanisms differ between Cepheids and AGB stars, the shared principle remains: the sustainability of overtone modes might be linked to the survival of radial nodes. Due to the dynamic interaction between convection and pulsations, these processes are difficult to disentangle and study in isolation. As a result, we can for now offer only qualitative, often speculative, insights into how convection affects overtone modes in AGB stars or other stars driven primarily by convective processes.

\subsection{Implications of non-radial pulsations}

Pulsation-induced shocks play a pivotal role in lifting material into cooler regions where dust forms, enabling radiation pressure to drive outflows \citep[e.g.][]{fleischerCircumstellarDustShells1992a,garcia-larioPropertiesPostAGBStars2006, gullieuszikVMCSurveyIII2012,liljegrenPulsationinducedAtmosphericDynamics2017b, hofnerDynamicAtmospheresWinds2022a}. The presence of multiple pulsation modes in AGB stars can give rise to non-spherical shocks in their extended atmospheres. These shocks, driven by the complex interplay between convection, radial and non-radial pulsations, disrupt the uniformity of the stellar envelope, leading to localised variations in density and temperature. Such inhomogeneities create conditions for episodic and spatially irregular dust formation, as the efficiency of dust condensation is highly sensitive to densities and temperatures \citep{hofnerExploringOriginClumpy2019, freytagGlobal3DRadiationhydrodynamical2023}. This process is further complicated by the coupling between convective flows and the pulsation modes, which can induce amplitude modulations and phase shifts, enhancing the irregularity of mass-loss episodes. The result is a circumstellar environment characterised by clumpy, asymmetric dust structures \citep{freytagGlobal3DRadiationhydrodynamical2023}. As dust-rich shells surrounding AGB stars significantly reprocess stellar radiation, these irregularities manifest in the spectral energy distribution and other observable signatures \citep[e.g.][]{piovanShellsDustAGB2003, maerckerDetachedDustShells2014, villaumeCircumstellarDustAGB2015, wiegertAsymmetriesAsymptoticGiant2024}, offering critical insights into the mass-loss processes and atmospheric dynamics of evolved stars.

The spatial asymmetries introduced by non-radial modes can create localised regions of compression and expansion, enhancing the formation of inhomogeneous shocks. The variability introduced by multi-mode pulsations, coupled with non-linear effects, could create fluctuations in mass-loss rates and lead to a diverse range of outcomes in circumstellar environments. These processes are critical for understanding the chemical enrichment of the interstellar medium, as AGB stars are major contributors to elements like carbon and nitrogen \citep[][]{tosiAGBStarsChemical2007, karakasHeavyelementYieldsAbundances2018}.

\section{Summary and conclusions}

We have investigated multi-mode pulsations in AGB stars using global 3D stellar models produced with the RHD code \texttt{CO5BOLD}. Our study explored the behaviour of radial and non-radial modes (up to the quadrupole mode) and different radial orders (up to the second overtone mode). We examined the transitions and interplay between these pulsation modes, as well as the influence of key stellar parameters, such as the mass and luminosity, which define the mean density and influence convection dynamics. A key strength of the \texttt{CO5BOLD} models lies in their fully self-consistent treatment of convection and pulsations in 3D, with no need for parameterisation of these processes, unlike traditional 1D models. This 3D approach allows for a more realistic representation of the stellar interior and atmosphere, capturing the non-linear interactions between pulsations and convection. These models offer predictive power for identifying and characterising pulsation modes that emerge naturally from the simulations.

Mode dominance in pulsating AGB stars evolves significantly as the stars progress through their evolutionary stages. The fundamental mode is predominant in luminous, large-radius stars with lower mean densities, while the first overtone mode dominates in denser, less luminous stars earlier in the AGB phase. This behaviour aligns closely with established P-L sequences and theoretical expectations. The derived pulsation periods generally follow the expected period-mean density relation, with deviations emerging primarily in transition regions where mode-switching occurs. Comparisons with Ritter's relation \citep{ritterUntersuchungenUeberHoehe1879} against linear fits to our models reveal  a good correlation between the pulsation period and the inverse square root of the mean density, particularly for the mode carrying the most significant power. The fundamental and first overtone modes were found to be dominant respectively below and above a mean density threshold of $\log(\bar{\rho}_\star \, [gcm^{-3}]) = -6.8$, reflecting dependence of pulsation modes to the stellar structure.

The persistence of pulsation nodes appears to depend critically on the interplay between convective and sound velocities. Regions with lower convective velocities relative to the local sound speed, particularly in high mean density stellar interiors, provide a stable environment for overtone modes. In contrast, vigorous convection in the outer layers of less dense stars may interfere with these modes, resulting in mode-switching behaviour. This dynamic interaction highlights the sensitivity of the amplitude of pulsations to the stellar structure and convective environment.

Using a wavelet transform (the WWZ method, in our case), we identified mode-switching and overtone activity in both our models and observational data. Comparisons with real stars: R Dor (a Mira-like star), RY Dra (an irregular semiregular variable), and AF Cyg (a low-amplitude overtone-dominated star), reveal qualitative agreement in pulsation mode behaviours and transitions. Observational data, particularly for semiregular variables, show active mode-switching, which can be primarily attributed to convection-driven variability. This analysis underscores the ability of our models to reproduce the complex pulsation dynamics observed in real AGB stars.

Finally, our models confirm that pulsation amplitudes are influenced by convection-pulsation interactions. Convective downdrafts and flow fluctuations, especially when synchronised with pulsation periods, could enhance or damp pulsation amplitudes. Such interactions may play a significant role in shaping mass-loss processes and the variability observed in AGB stars. This behaviour warrants further investigation, as it could provide deeper insights into the mechanisms driving episodic mass loss and the evolution of circumstellar environments.

\section*{Data availability}
An electronic table, consisting of global stellar parameters of all models together with the derived pulsation periods analysed in this work is available in electronic form at the CDS via anonymous ftp to \href{https://cdsarc.cds.unistra.fr/}{cdsarc.u-strasbg.fr} (\href{ftp://130.79.128.5/}{130.79.128.5}) or via \href{http://cdsweb.u-strasbg.fr/cgi-bin/qcat?J/A+A/}{http://cdsweb.u-strasbg.fr/cgi-bin/qcat?J/A+A/}.

\begin{acknowledgements}
      This work is part of a project that has received funding from the European Research Council (ERC) under the European Union's Horizon 2020 research and innovation programme (Grant agreement No. 883867, project EXWINGS) and the Swedish Research Council (Vetenskapsrådet, grant number 2019-04059). The computations were performed on resources provided by the Swedish National Infrastructure for Computing (SNIC) and the National Academic Infrastructure for Supercomputing in Sweden (NAISS), supported by the Swedish Research Council (Vetenskapsrådet).
\end{acknowledgements}

%
%

\bibliographystyle{aa_url}
\bibliography{refs}

\label{LastPage}
\end{document}